# Phase diagram of a strongly disordered s-wave superconductor, NbN, close to the metal-insulator transition

Madhavi Chand[a], Garima Saraswat[a*], Anand Kamlapure[a†], Mintu Mondal[a], Sanjeev Kumar[a], John Jesudasan[a], Vivas Bagwe[a], Lara Benfatto[b], Vikram Tripathi[a] and Pratap Raychaudhuri[a‡]

[a]Tata Institute of Fundamental Research, Homi Bhabha Road, Colaba, Mumbai 400005, India.
[b]ISC-CNR and Department of Physics, Sapienza University, Piazzale Aldo Moro 5, 00185 Rome, Italy.

*Abstract:* We present a phase diagram as a function of disorder in three-dimensional NbN thin films, as the system enters the critical disorder for the destruction of the superconducting state. The superconducting state is investigated using a combination of magnetotransport and tunneling spectroscopy measurements. Our studies reveal 3 different disorder regimes. At low disorder ($k_F l$~10-4), the system follows the mean field Bardeen-Cooper-Schrieffer behavior where the superconducting energy gap vanishes at the temperature where electrical resistance appears. For stronger disorder ($k_F l$<4) a "pseudogap" state emerges where a gap in the electronic spectrum persists up to temperatures much higher than $T_c$, suggesting that Cooper pairs continue to exist in the system even after the zero resistance state is destroyed. Finally, at even stronger disorder ($k_F l$<1) the global superconducting ground state is completely destroyed, though superconducting correlations continue to survive as evidenced from a pronounced magnetoresistance peak at low temperatures.

---

[*] E-mail: gary@tifr.res.in
[†] E-mail: ask@tifr.res.in
[‡] E-mail: pratap@tifr.res.in

# I. Introduction

In recent years, the effect of strong disorder in conventional s-wave superconductors has attracted renewed attention, motivated by the observation of novel electronic phases close to the critical disorder where superconductivity gets destroyed. In the low disorder limit, based on Bardeen-Cooper-Schrieffer (BCS) theory, Anderson[1] postulated that the superconducting transition temperature ($T_c$) of a superconductor will remain unchanged. However, subsequent measurements on a wide variety of systems[2,3,4,5,6,7,8] showed that as the disorder level is increased towards the strong disorder limit, $T_c$ gradually decreases, eventually leading to a non-superconducting ground state. It is now understood that superconducting correlations continue to play a dominant role in the electronic properties even after the global superconducting ground state is completely destroyed. These correlations manifest through several phenomena: A giant peak in the magnetoresistance in strongly disordered superconducting films[9,10,11,12,13], the persistence of magnetic flux quantization in strongly disordered Bi films even after the film is driven into an insulating state[14], finite high-frequency superfluid stiffness above the superconducting transition temperature[15], and more recently, the observation of a pronounced "pseudogap" in the electronic spectra of several strongly disordered superconductors[16,17,18] which persists up to temperatures many times $T_c$. These observations lead to the obvious question on whether strong disorder can destroy the superconducting state without suppressing the underlying pairing interactions, leading to electronic states with finite Cooper pair density but no global superconductivity.

The suppression of superconductivity in the presence of strong disorder is driven by two distinct, but not mutually exclusive effects. The first effect results from the increase in electron-electron (*e-e*) Coulomb repulsion caused by a loss of effective screening[19,20], which partially cancels the electron-phonon mediated attractive pairing interaction. The second effect comes from the decrease in superfluid density ($n_s$) induced by disorder scattering[21] in the presence of strong disorder. Since reduced $n_s$ and the loss of effective screening, both render the superconductor susceptible to quantum and classical phase fluctuations[22], enhanced phase fluctuations can destroy the superconducting state even when the pairing amplitude remains finite[23]. While both these mechanisms have been invoked to explain different sets of experimental observations, a hierarchical scheme to understand the relative importance of these effects at different levels of disorder is at present lacking.

In this paper, we address this issue through a combination of magnetotransport and tunneling measurements on three-dimensional NbN thin films (with thickness much larger than the dirty limit coherence length, $\xi$) grown through reactive magnetron sputtering. The disorder is tuned by controlling the level of Nb vacancies in the lattice, which is controlled by changing the Nb/N ratio in the plasma. For each film, we characterize the effective disorder through $k_Fl$, where $l$ is the electronic mean free path and $k_F$ is the Fermi wave vector. The disorder in this set of films spans a wide range, from the moderately clean limit ($k_Fl$~10) down to $k_Fl$~0.4. Consequently, the normal state resistivity ($\rho$) at low temperatures varies by 5 orders of magnitude, and $T_c$ ranges from 17K in the cleanest sample to <300mK for the samples with $k_Fl$<1. Our study reveals three different regimes: At moderate disorder ($10 > k_Fl > 4$), $T_c$ starts getting gradually suppressed but the system continues to follow conventional BCS behavior where the superconducting energy gap disappears at the temperature where resistance appears. For stronger disorder ($4 > k_Fl > 1$), a "pseudogap" state emerges where a gap in the electronic energy spectrum persists up to a temperature $T^* >> T_c$. Finally at even stronger disorder ($k_Fl < 1$) we obtain a non-superconducting state, characterized by a pronounced peak in the magnetoresistance at low temperatures. Based on these observations we construct a phase diagram which clearly delineates the relative importance of different mechanisms at different levels of disorder.

## II. Experimental Details

Epitaxial thin films of NbN were grown using reactive magnetron sputtering on (100) oriented single crystalline MgO substrates, by sputtering a Nb target in Ar/N$_2$ gas mixture. The thickness of all our films, measured using a stylus profilometer was ≳50nm, which is much larger than the dirty-limit coherence length[24] ($\xi$ ~ 4-8 nm) in the superconducting state. Thus from the point of view of superconducting fluctuations all our samples are effectively in the 3-dimensional limit. The effective disorder, resulting from the amount of Nb vacancies in the NbN crystalline lattice, was controlled by controlling the sputtering power and/or the Ar/N$_2$ gas mixture both of which effectively changed the Nb/N ratio in the plasma. Details of synthesis and structural characterization of the films have been reported elsewhere[25,26,27].

Resistance ($R$), magnetoresistance (MR=($\rho(H)$-$\rho(0)$)/$\rho(0)$) and Hall effect measurements were performed using standard four-probe techniques from 285K down to 300mK using either a conventional $^4$He or $^3$He cryostat up to a maximum field of 12T. For each film, $k_Fl$ was determined from $\rho$ and Hall coefficient ($R_H$) measured at 285K using the free electron formula, $k_Fl = \left\{(3\pi^2)^{2/3}\hbar[R_H(285K)]^{1/3}\right\}/\left[\rho(285K)e^{5/3}\right]$, where $\hbar$ is Plank's constant and $e$ is the electronic charge. $R_H$ was calculated from the Hall voltage deduced from reversed field sweeps from 12T to -12T after subtracting the resistive contribution. We would like to note that while in a non-interacting scenario $k_Fl$ could provide a unique measure of electronic disorder, $e$-$e$ interactions can significantly alter this scenario. We therefore calculate $k_Fl$ from $R_H$ and $\rho$ measured at the highest temperature of our measurements (285K) where the effect of $e$-$e$ interactions is expected to be small[28]. The upper critical field ($H_{c2}$) for several samples was measured from either $\rho(T)$–$T$ scans at different magnetic fields ($H$) or $\rho(H)$–$H$ scans at different temperatures (with $H$ perpendicular to the plane of the film).

Scanning tunneling spectroscopy (STS) measurements were performed using a home-built, high-vacuum, low-temperature scanning tunneling microscope[29] (STM) operating down to 2.6K. The samples used in STS measurements were grown *in-situ*, in a sputtering chamber connected to the STM. A pair of horizontal and vertical manipulators was used to transfer the sample from the growth chamber to the STM without exposing to air. The tunneling density of states (DOS) was extracted at various temperatures, from the measurement of tunneling conductance ($G(V) = \left(\dfrac{dI}{dV}\right)$) as a function of voltage (*V*) between the sample and a Pt-Ir tip using a lock-in based voltage modulation technique operating at 312Hz and a modulation voltage of 100μV.

### III. Results

We first summarize the evolution of the zero field transport properties with disorder. Figure 1(a) shows $\rho$-$T$ for NbN films with $k_F l$ ranging from 10.12 to 0.42. All samples, other than the one with $k_F l$~10.12 show a negative temperature coefficient[26] of $\rho$. For samples with $k_F l$>1, $T_c$ (defined as the temperature where resistance reaches 1% of its normal state value) varies from 16K to <300 mK with increasing disorder. The samples with $k_F l$<1 remain non-superconducting down to 300 mK. Figure 1(b) shows the variation of $T_c$ with $k_F l$. We observe that $T_c \rightarrow 0$ as $k_F l \rightarrow 1$. The carrier density (*n*) at 285K ($n=1/(eR_H(285K))$) and the normal state resistivity ($\rho(285K)$) for all samples are shown in Figure 1(c). In the same graph we also plot the maximum resistivity, $\rho_m$ (taken as the peak value of $\rho$ before the onset of the superconducting transition for $k_Fl>1$ and $\rho$ at 300mK for $k_Fl<1$) which varies by 5 orders of magnitude from 0.5 $\mu\Omega$ m to 15000 $\mu\Omega$ m over the entire range of disorder. In this context we would like to note that while $k_Fl\sim 1$ is usually associated with the Anderson metal-insulator transition, such a classification is not straightforward in a disordered electronic system where electron-electron interactions can play a significant role. For our films with $k_Fl<1$, the conductivity ($\sigma=1/\rho$) increases linearly with temperature (Figure 1(d)) and shows a small positive intercept when extrapolated to T$\rightarrow$0, typical of a "bad" metal[30]. In our opinion, the bad metallic behavior observed for $k_Fl<1$ reflects the inaccuracy associated with the determination of this parameter based on free-electron formula, in a system where $e$-$e$ interactions arising from the diffusive motion of the electrons could be important. We have observed remarkable consistency between the $\rho_m$, $T_c$ and $n$ for different films grown over a period of more than two years

In order to explore the superconducting state, STS measurements were performed on several films with different levels of disorder. The measurement was performed by recording $G(V)$ vs. $V$ on 32 equally spaced points along a 150nm line at different temperatures, which allowed us to obtain information on both the temperature evolution as well as the spatial variation of the tunneling DOS in the sample. Figure 2(a-f) shows the normalized conductance, $G(V)/G_N$ (where $G_N=G(V\gg \Delta/e)$) as a function of $V$, averaged over the 32 points, at different temperatures for 6 samples with different disorder. For the first three samples (Fig. 2(a-c)) where the lowest temperature of measurement is less than $T_c$, the conductance spectra show a dip at V=0 and two symmetric peaks as expected from BCS theory. This feature is also observed in the next two disordered samples (Fig. 2(d-e)) where the lowest temperature of our measurements is higher than $T_c$. The most disordered sample ($T_c\ll$300 mK) shows a dip in $G(V)$ for $V\lesssim 2$ mV which rides over a broader "V" shaped background which extends up to high bias. This nearly temperature independent broad background, is observed for all our samples and persists up to the highest temperature of our measurements. It arises from Altshuler-Aronov type *electron-electron* interactions and becomes more pronounced for samples with higher disorder[31]. In order to isolate the feature associated with superconductivity, for each sample, we subtract this background using the spectra at the temperature above which the low-bias feature associated with superconductivity disappears (shown as thick lines). The background corrected conductance spectra ($G_{sub}(V)$ *vs.* *V*), normalized at high bias, are shown in Figure 2(g-l). In the $G_{sub}(V)$ *vs.* *V* spectra, the broadened coherence peaks are visible in all the samples. In Fig. 3(a-f) we plot the temperature evolution of $G_{sub}(V)$ in the form of an intensity plot (averaged over 32 points as in Figure 2), as a function of temperature and bias voltage, for films with different disorder. The lower panel of each plot shows the *R-T* measured on the same films. For the most ordered film ($T_c$~11.9K), the features in the tunneling DOS associated with superconductivity disappear at $T_c$, thereby restoring a flat metallic DOS for $T>T_c$. However with increase in disorder, the low-bias dip in $G_{sub}(V)$ *vs.* *V* spectra continues to persist up to a characteristic temperature $T^*>T_c$. It is interesting to note that the pseudogap temperature ($T^*$) remains almost constant for samples with $T_c \lesssim 6K$. Figure 4(a-f) shows the spatial variation of $G_{sub}(V)$ recorded at the lowest temperature along a 150 nm line for each sample. While the zero bias dip and the two symmetric peaks are uniform over the entire line for the sample with $T_c$~11.9K, the superconducting state becomes progressively inhomogeneous with increase in disorder. For the two most disordered samples, for which $T_c$ is smaller than the base temperature of our STM, the local DOS in the pseudogap state shows superconducting domains, few tens of nanometers in size, separated by regions where the superconducting feature is completely suppressed. A similar situation is also observed in other samples in the temperature range $T_c<T<T^*$. This is shown in Figure 5 where we show the spatial variation of $G_{sub}(V)$ at different temperatures for a sample with $T_c$~2.7K.

Finally, we focus our attention on the magnetotransport properties in the strong disorder limit. Figure 6(a) shows $\rho$-$H$ at different temperatures for the most disordered film with $k_Fl$~0.42. $\rho$-$H$ shows a pronounced peak at a characteristic field ($H_p$) which gradually disappears with increase in temperature. In the most disordered sample the resistance at 12T is smaller than the corresponding zero field value. This peak becomes less pronounced (Fig. 6(b-c)) as the disorder is reduced and completely disappears for films with $k_Fl \gtrsim 1$. It is interesting to note that the peak in $\rho$-$H$ disappears at a temperature which is close to $T^*$ for the most disordered sample on which STS was performed (Fig. 6 (e)-(f)). For the film with $k_Fl$~1.23 which has $T_c$~0.6K (Fig. 6(d)), at 300mK, $\rho$ increases monotonically with $H$, exhibiting a broad transition to the normal state as expected for a strongly disordered sample. As expected, for this sample, a positive MR is observed even at $T>>T_c$ originating from superconducting fluctuations which persist above $T_c$.

Figure 7 summarizes the evolution of the superconducting state as a function of disorder in the form of a phase diagram, where we plot $T_c$ and $T^*$ as a function of $k_Fl$. This phase diagram brings out three distinct regimes of disorder: (i) The intermediate disorder regime (marked as **I**), where the superconducting state is characterized by a single energy scale, $T_c$; (ii) the strongly disordered regime (marked as **II**), which is characterized by the emergence of a second energy scale, $T^*>T_c$, up to which the superconducting energy gap persists in the tunneling spectrum; and (iii) an even more strongly disordered regime (marked as **III**) ($k_Fl < 1$), which does not exhibit any superconducting transition, but exhibits a pronounced peak in the MR which disappears at temperatures close to $T^*$. In the next section we will discuss the various mechanisms that contribute to these behaviors.

**IV. Discussion**

Before discussing how different energy scales emerge in a superconductor with increase in disorder, we briefly summarize the mechanisms responsible for the destruction of superconductivity. The superconducting state is characterized by a complex order parameter $\Psi = |\Delta|e^{i\phi}$, where $|\Delta|$ is a measure of the binding energy of the Cooper pairs and $\phi$ is the phase of the macroscopic condensate. It is important to note that a finite $|\Delta|$ manifests as a gap in the electronic energy spectrum, whereas the zero resistance state results from the phase coherence of the Cooper pairs over all length scales. The first route, through which superconductivity can get suppressed, is by a decrease in $|\Delta|$ caused by a weakening of the pairing interactions. In such a situation, $T_c$ will get suppressed but the superconductor will continue to follow conventional BCS behavior with the superconducting energy gap disappearing at $T_c$. However, a second, less explored route for the suppression of $T_c$ is through a decrease in the phase stiffness[22,32]. When the phase stiffness becomes sufficiently small the superconducting state will get destroyed due to a loss of global phase coherence resulting from thermally excited phase fluctuations, leaving pairing amplitude ($|\Delta|$) finite above $T_c$. In such a situation the superconducting energy gap will continue to persist for $T \gg T_c$, till a temperature is reached where the pairing amplitude also vanishes.

In region **I** of the phase diagram, $T_c$ monotonically decreases with increase in disorder, but continues to follow conventional BCS behavior. Therefore, we expect the decrease in $T_c$ to be caused by a weakening of the pairing interaction. This weakening can result from two effects. First, with increase in disorder, the diffusive motion of the electron results in an increase in the repulsive *e-e* Coulomb interactions[19], which partially cancels the phonon mediated attractive pairing interaction. It is interesting to note that some of the early works attributed the complete suppression of superconductivity in several disordered superconductors[5,6], solely to this effect[19,20]. The second effect comes from the fact that disorder, in addition to localizing the electronic states close to the edge of the band also increases the one electron bandwidth[33], thereby decreasing the density of states ($N(0)$) close the middle of the band. While this effect alone cannot result in complete suppression of superconductivity, it can have a noticeable effect in the intermediate disordered regime[34]. Both these effects are captured at a qualitative level using the modified BCS relation[35], $T_c = 1.13\Theta_D \exp\left(-\dfrac{1}{N(0)V - \mu^*}\right)$, where $\Theta_D$ is a temperature scale of the order of Debye temperature, $V$ is the attractive electron-phonon potential and $\mu^*$ is the Coulomb pseudopotential which accounts for the disorder enhanced $e$-$e$ interactions. While the available theoretical model on the dependence of the $\mu^*$ on disorder in a 3-D superconductor is currently not developed enough to attempt a quantitative fit of our data, the combination of the two effects mentioned above qualitatively explains the suppression of $T_c$ in region **I,** where the superconducting energy gap in the tunneling DOS vanishes exactly at $T_c$.

As the disorder is further increased, the superconductor enters regime **II,** which is characterized by two temperature scales, namely, $T_c$, which corresponds to the temperature at which the resistance appears and $T^*$, which corresponds to the temperature at which the superconducting energy gap disappears. $T_c$ continues to decrease monotonically with increasing disorder, whereas $T^*$ remains almost constant[36] down to $k_Fl$~1, where the superconducting ground state is completely destroyed. It would be natural to ascribe these two temperature scales to the phase stiffness of the superfluid ($J$) and the strength of the pairing interaction ($|\Delta|$) respectively. $J$ can be estimated using the relation[22],

$$J = (\hbar^2 a n_s)/(4m^*), \qquad (1)$$

where $a$ is the length scale over which the phase fluctuates and $m^*$ is the effective mass of the electron. A rough estimate of $J$ is obtained from $n_s$ derived from the low temperature penetration depth[16] ($\lambda(T\rightarrow 0)$) and setting $a \approx \xi$. In conventional "clean" superconductors, $J$ is several orders of magnitude larger than $|\Delta|$, and therefore phase fluctuations play a negligible role in determining $T_c$. However, disorder enhanced electronic scattering decreases $n_s$, thereby rendering a strongly disordered superconductor susceptible to phase fluctuations. In Figure 8, we summarize the values of $J$ for NbN films with different $T_c$ estimated from eqn. (1) using experimental values of $n_s$ measured from penetration depth (ref. 16) and the values of $\xi$ obtained from the upper critical field, $H_{c2}$ (ref.24)). Apart from some small numerical factor of the order of one arising from the choice of the cut-off $a \approx \xi$ in eqn. (1), we see that while for the samples in regime **I**, $J \gg k_B T_c$ such that phase fluctuations are irrelevant, as we enter regime **II**, $J \lesssim k_B T_c$. Moreover, the crossover from regime **I** to regime **II** occurs on the same samples where we observe a deviation of $n_S(T)$ from the dirty-limit BCS theory, both at zero temperature and finite $T$ (Ref. 16). Both effects can be attributed to phase fluctuations in the presence of disorder. As it has been recently discussed in Ref. 37, as disorder increases, the superfluid stiffness is lower than in the dirty-BCS scenario since the phase of the superconducting order parameter relaxes to accommodate to the local disorder, leading to an additional paramagnetic reduction of the superfluid response of the system. At the same time the enhanced dissipation lowers the temperature scale where longitudinal phase fluctuations can be excited, leading to a linear decrease of $n_s(T)$ in temperature, as observed in our samples[16]. In light of these observations, we therefore conclude that the superconducting state in strongly disordered NbN samples is destroyed at $T_c$ due to phase fluctuations between superconducting domains that are seen to spontaneously form in our STS data (Fig.4 and Fig.5). However, even above this temperature, $|\Delta|$ remains finite due to phase incoherent Cooper pairs which continue to exist in these domains. The relative insensitivity of $T^*$ to disorder and the gradual decrease in $T_c$ suggests that increase in phase fluctuations is responsible for the decrease in $T_c$ in this regime, while the pairing amplitude remains almost constant. Eventually, at a critical disorder ($k_Fl \approx 1$), the superconducting ground state is completely suppressed by quantum phase fluctuations, that are themselves enhanced by disorder. The overall physical picture and the phase diagram obtained in our experiments share many analogies with recent theoretical calculations on disordered superconductors[34,37,38].

As the disorder in increased further, we enter regime **III**, where all samples remain non-superconducting down to 300 mK. This phase is characterized by a peak in the MR which is a hallmark of several strongly disordered superconductors[9,10,11,12,13]. Since the pairing amplitude remains finite down to the critical disorder where $T_c \rightarrow 0$, it is expected that superconducting correlations will continue to play a significant role in this regime. The superconducting origin of the MR peak is suggested from the fact that it vanishes at temperatures close to $T^*$ measured from STS in samples in regime **II**. Numerical simulations[39] (in 2D) also indicate that the non-superconducting state could comprise of small superconducting islands, where quantum phase fluctuations between these islands prevent the establishment of global superconducting order. It has been shown that such an inhomogeneous scenario[40,41,42] can give rise to the non-monotonic variation of $\rho$-$H$. At low fields, the increase in $\rho$ reflects the gradual decrease in superconducting paths for the current to flow due to the shrinkage in the size of the superconducting droplets. However, at high fields when the superconducting islands become very small the superconducting regions are avoided by the current and the decrease in resistance is caused by the gradual increase in normal regions through which the current flows[43]. In such a scenario, $H_p$ is associated with the mean value of magnetic field where superconducting correlations are almost destroyed in the sample. $H_p$ is therefore expected to evolve smoothly from $H_{c2}$ in the superconducting state as one enters Regime **III** from Regime **II**. To verify this we compare $H_p$ measured at 300 mK for samples with $k_Fl<1$ with $H_{c2}(0)$ for samples with $k_Fl>1$. For the samples with $T_c<5K$ $H_{c2}(0)$ is determined from $\rho$-$H$ scans, at the field where $\rho$ reaches 90% of its normal state value at the lowest temperature of our measurements. For films with higher $T_c$, $H_{c2}(0)$ was estimated from the temperature variation of $H_{c2}(T)$ close to $T_c$, using the dirty-limit formula[44], $H_{c2}(0) = 0.69 T_c (dH_{c2}/dT)\big|_{T=T_c}$. Figure 9 shows the evolution of $H_{c2}(0)$ and $H_p$ as a function of $k_Fl$. We observe that with increasing disorder $H_{c2}(0)$ monotonically decreases and smoothly connects to $H_p$ for the samples in regime **III,** providing a further confirmation of the superconducting origin of the MR peak**.**

## IV. Summary

To summarize, we have shown how with increase in disorder, a 3D conventional superconductor, NbN, evolves from a BCS superconductor in the moderately clean limit, to a situation where the destruction of the superconducting state is governed by strong phase fluctuations. Based on transport and STS measurements on 3D films spanning a large range of disorder, we construct a phase diagram where we can identify the dominant interactions in different regimes of disorder: (i) The intermediate disorder regime, where $T_c$ decreases due to a gradual weakening of the pairing interaction; (ii) a strongly disordered regime, where $T_c$ is governed by a decrease in the superfluid stiffness, though the pairing strength remains almost constant; and (iii) a non-superconducting ground state at even stronger disorder formed of phase incoherent superconducting puddles/islands. It would be worthwhile to carry out similar measurements on other strongly disordered superconductors such as $InO_x$ or TiN to explore the extent to which such a phase diagram is generic for all disordered s-wave superconductors. It would also be interesting to explore to what extent such a scenario could be applicable to underdoped high-$T_c$ cuprates, which share many similarities with strongly disordered s-wave superconductors.

References:

[1] P. W. Anderson, J. Phys. Chem. Solids **11**, 26 (1959).

[2] A. M. Goldman and N. Markovic, Phys. Today **51,** No. 11, 39 (1998).

[3] A. T. Fiory and A. F. Hebard, Phys. Rev. Lett. **52,** 2057 (1984).

[4] R. C. Dynes, J. P. Garno, G. B. Hertel, and T. P. Orlando, Phys. Rev. Lett. **53**, 2437 (1984).

[5] T. Furubayashi, N. Nishida, M. Yamaguchi, K. Morigaki, H. Ishimoto, Solid State Commun. **55**, 513 (1985).

[6] G. Hertel, D. J. Bishop, E. G. Spencer, J. M. Rowell, and R. C. Dynes, Phys. Rev. Lett. **50**, 743 (1983).

[7] J. Lesueur, L. Dumoulin and P. Nedellec, Solid State Commun. **66,** 723 (1988).

[8] K. E. Gray, R. T. Kampwirth, T. F. Rosenbaum, S. B. Field and K. A. Muttalib, Phys. Rev. B **35,** 8405 (1987).

[9] V. F. Gantmakher, M. V. Golubkov, V. T. Dolgopolov, G. E. Tsydynzhapov and A. A. Shashkin, JETP Letters **68,** 363 (1998).

[10] M. Steiner and A. Kapitulnik, Physica C **422**, 16 (2005).

[11] G. Sambandamurthy, L. W. Engel, A. Johansson, E. Peled, and D. Shahar, Phys. Rev. Lett. **94**, 017003 (2005).

[12] T. I. Baturina, C. Strunk, M. R. Baklanov and A. Satta, Phys. Rev. Lett. **98**, 127003 (2007).

[13] H. Q. Nguyen, S. M. Hollen, M. D. Stewart, Jr., J. Shainline, Aijun Yin, J. M. Xu and J. M. Valles, Jr., Phys. Rev. Lett. **103**, 157001 (2009).

[14] M. D. Stewart Jr., A. Yin, J. M. Xu and J. M. Valles Jr., Science **318,** 1273 (2007).

[15] R. Crane, N. P. Armitage, A. Johansson, G. Sambandamurthy, D. Shahar, and G. Grüner, Phys. Rev. B **75,** 184530 (2007).

[16] M. Mondal, A. Kamlapure, M. Chand, G. Saraswat, S. Kumar, J. Jesudasan, L. Benfatto, V. Tripathi and P. Raychaudhuri, Phys. Rev. Lett. **106**, 047001 (2011).

[17] B. Sacépé, C. Chapelier, T. I. Baturina, V. M.Vinokur, M. R. Baklanov and M. Sanquer, Nat Commun. **1**, 140 (2010).

[18] B. Sacépé, T. Dubouchet, C. Chapelier, M. Sanquer, M. Ovadia, D. Shahar, M. Feigel'man and L. Ioffe, Nature Physics **7,** 239–244 (2011).

[19] P. W. Anderson, K. A. Muttalib and T. V. Ramakrishnan, Phys. Rev. B **28**, 117 (1983).

[20] A. M. Finkelstein, Physica B **197**, 636-648 (1994).

[21] M. Tinkham, *Introduction to Superconductivity* (McGraw-Hill, New York, 1996).

[22] V. J. Emery and S. A. Kivelson, Nature (London) **374**, 434 (1995); Phys. Rev. Lett. **74**, 3253 (1995).

[23] It has also been proposed that when the system is driven into an insulating state by strong disorder, the Cooper pairs can themselves get localized over short length scales, giving rise to an insulating state consisting of localized Cooper pairs: M. V. Feigel'man, L. B. Ioffe, V. E. Kravtsov, and E. A. Yuzbashyan, Phys. Rev. Lett. **98**, 027001 (2007); Ann. Phys. (N.Y.) **325,** 1390 (2010). This is however beyond the disordered regime explored in this paper.

[24] M. Mondal, M. Chand, A. Kamlapure, J. Jesudasan, V. C. Bagwe, S. Kumar, G. Saraswat, V. Tripathi and P. Raychaudhuri, J. Supercond. Nov. Magn. **24,** 341 (2011).

[25] S. P. Chockalingam, M. Chand, J. Jesudasan, V. Tripathi and P. Raychaudhuri, Phys. Rev. B **77**, 214503 (2008).

[26] M. Chand, A. Mishra, Y. M. Xiong, A Kamlapure, S. P. Chockalingam, J. Jesudasan, V. Bagwe, M. Mondal, P. W. Adams, V. Tripathi and P. Raychaudhuri, Phys. Rev. B **80**, 134514 (2009).

[27] S. P. Chockalingam, M. Chand, J. Jesudasan, V. Tripathi and P. Raychaudhuri, J. Phys.: Conf. Ser. **150**, 052035 (2009).

[28] M. A. Khodas and A. M. Finkelstein, Phys. Rev. B **68,** 155114 (2003).

[29] Details of our STM are given in Ref. 7.

[30] E. Tousson and Z. Ovadyahu, Solid State Commun. **60,** 407 (1986); A. T. Fiory and A. F. Hebard, Phys. Rev. Lett. **52**, 2057 (1984).

[31] For example $G(V=3.7$ mV$)/G(0)$ just above $T^*$ increases monotonically from 1.04 to 1.17 as we go across this series from the least disordered sample to the most disordered one.

[32] T. V. Ramakrishnan, Phys. Scr., T **27,** 24 (1989).

[33] J. M. Ziman, *Models of Disorder,* (Cambridge University Press, 1979).

[34] A. Ghosal, M. Randeria, and N. Trivedi, Phys. Rev. B **65**, 014501 (2001).

[35] W. L. McMillan, Phys. Rev. **167**, 331 (1968).

[36] Since *e-e* interactions continue to increase with disorder in Regime **II** (for example, evidenced from a gradual increase in $G(V=3.7$ mV$)/G(0)$ (ref. 31) at temperature just above $T^*$) the relative insensitivity of $T^*$ with disorder for samples with $T_c$<6 K is somewhat counterintuitive. While we do not have an explanation for this behavior at present, we would like to note that in the absence of *e-e* interactions, a gradual increase in the spectral gap and its associated temperature scale at strong disorder has been found in the numerical work of refs. 34 and 39.

[37] G. Seibold, L. Benfatto, C. Castellani and J. Lorenzana, (arXiv:1107.3839, unpublished).

[38] X. T. Wu and R. Ikeda, Phys. Rev. B **83**, 104517 (2011).

[39] K. Bouadim, Y. L. Loh, M. Randeria, N. Trivedi, Nature Phys. **7**, 884 (2011).

[40] Y. Dubi, Y. Meir and Y. Avishai, Phys. Rev. B **73,** 054509 (2006).

[41] For a 3-D superconductor-insulator transition deep in the insulating regime, it has been suggested (Ref. 23) that the insulating and superconducting state could both comprise of Cooper pairs which are delocalized on the superconducting side and localized on the insulating side of the transition. While such a scenario could also give a MR peak at the pair breaking field (Ref. 42), it is unlikely to be applicable in our samples which are not deep in the insulating regime.

[42] M. Mueller, (arxiv: 1109.0245, unpublished)

[43] A similar mechanism for the decrease in resistance at high fields has been proposed in the context of granular superconductors, I. S. Beloborodov, K. B. Efetov, A. I. Larkin, Phys. Rev. B **61,** 9145 (2000).

[44] N. R. Werthamer, E. Helfland, and P. C. Honenberg, Phys. Rev. **147**, 295 (1966).

**Figures:**

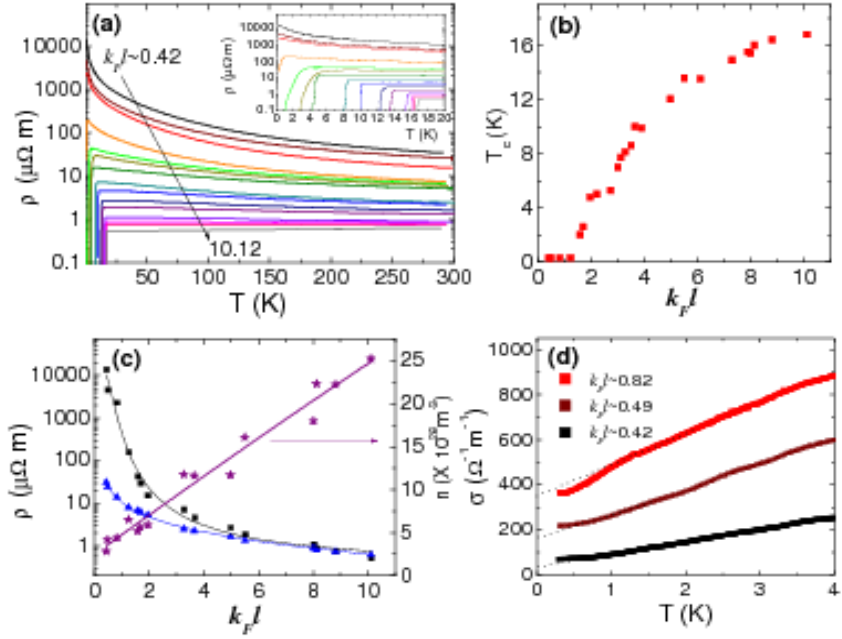

**Figure 1.** (a) $\rho$ vs. $T$ for NbN films with different $k_Fl$; the *inset* shows the expanded view at low temperatures. (b) Variation of $T_c$ with $k_Fl$. (c) Variation of $n$ (★), $\rho(285K)$ (▲) and $\rho_m$ (■) with $k_Fl$. (d) Conductivity ($\sigma$) vs. $T$ at low temperature for the samples with $k_Fl \sim$ 0.82, 0.49 and 0.42. The dotted lines are extrapolations to $\sigma(T\rightarrow 0)$.

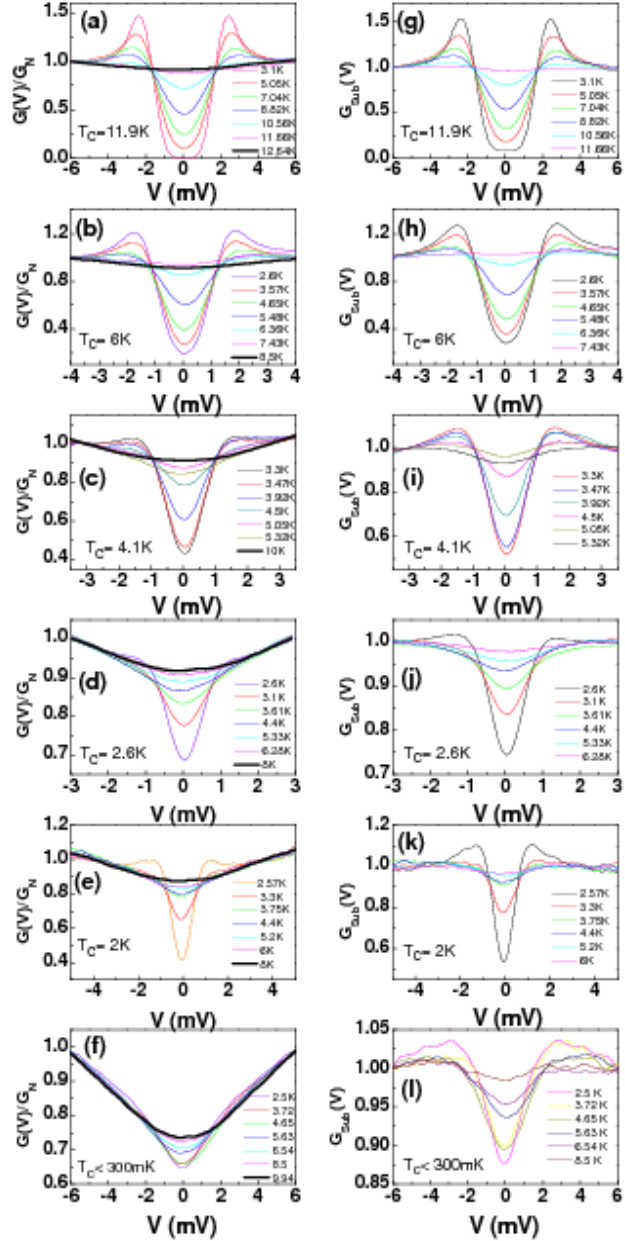

**Figure 2.** (a)-(f) Normalized tunneling spectra at different temperatures for NbN films with different disorder. The spectrum shown in thick line corresponds to the temperature at which the low bias feature in the tunneling conductance disappears. Each tunneling spectrum is averaged over 32 equally spaced points along a 150nm line on the sample surface. (g)-(l) The spectra corresponding to (a-f) after subtracting the V shaped background.

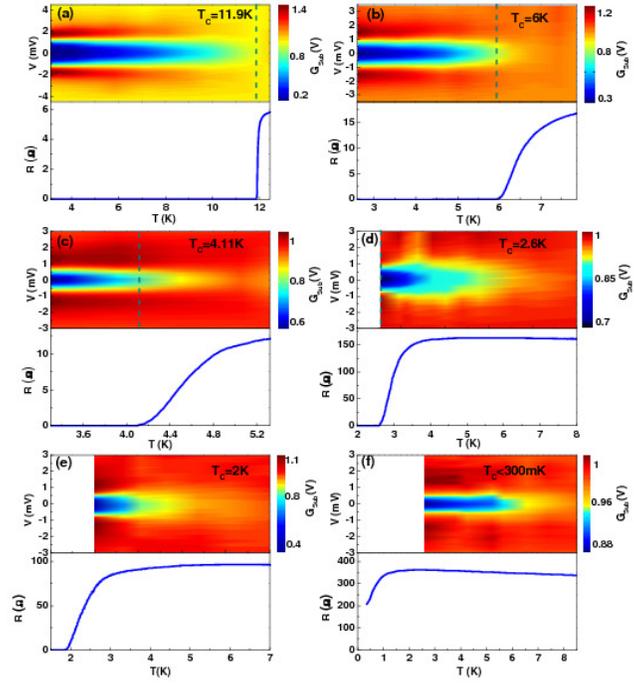

**Figure 3.** (a)-(f) Intensity plot of $G_{sub}(V)$ as a function of temperature and applied bias for 6 different samples (upper panels) along with resistance versus temperature in the same temperature range (lower panels).

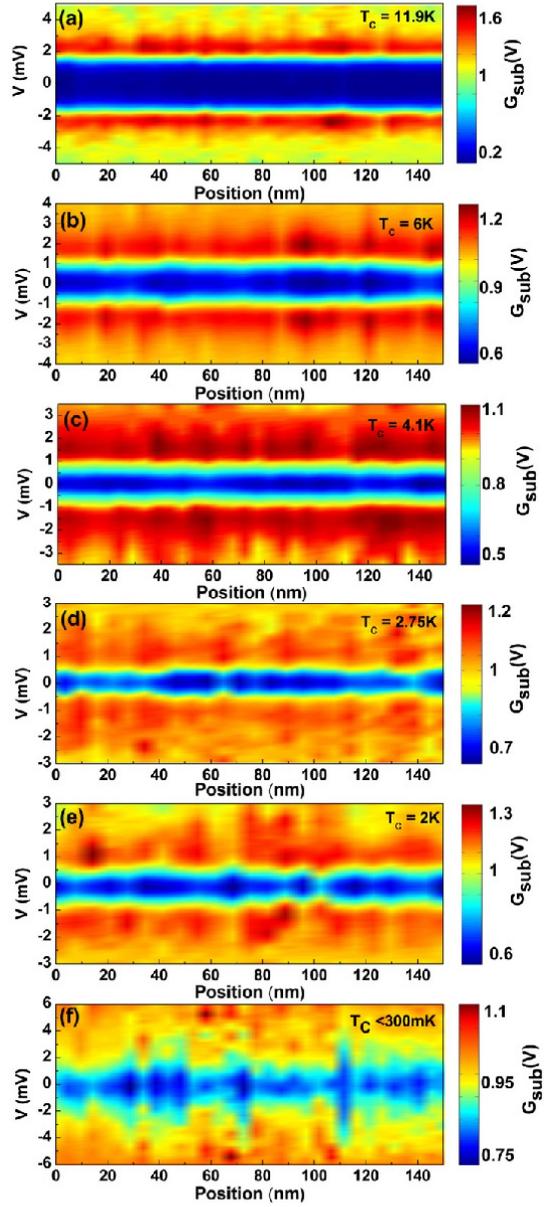

**Figure 4.** (a)-(f) $G_{sub}(V)$ *vs.* $V$ spectra along a 150nm line (measured at 2.6K) for six NbN thin films with different disorder.

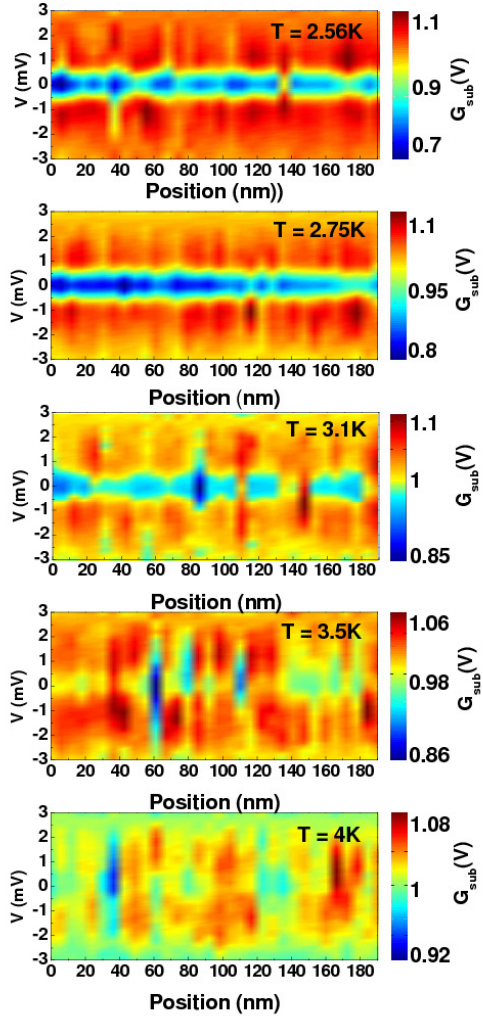

**Figure 5.** Spatial variation of $G_{sub}(V)$ *vs.* $V$ spectra recorded along a 190 nm line at different temperatures for an NbN thin film with $T_c$~2.7K. Large inhomogeneity in the tunneling DOS is observed as we enter the pseudogap state.

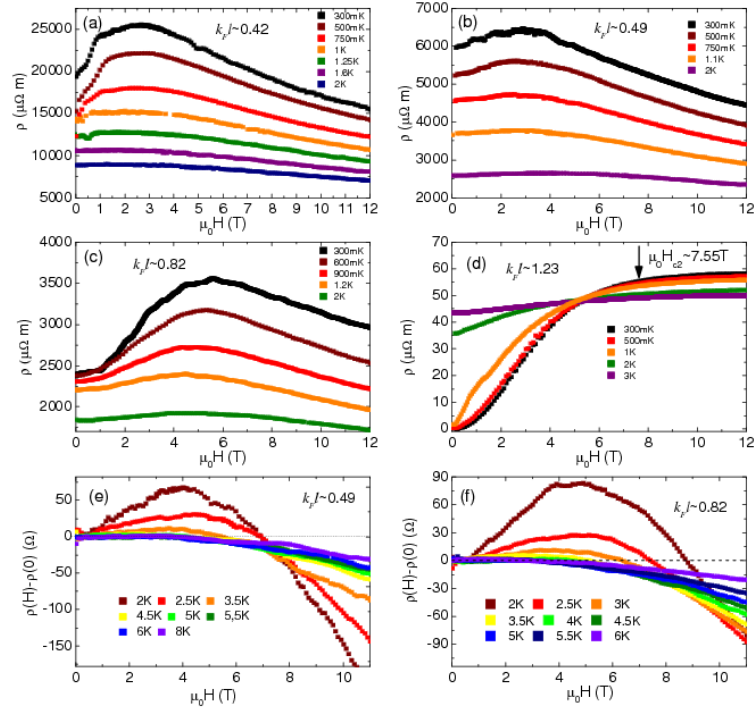

**Figure 6.** Resistivity as a function of magnetic field at different temperatures for 4 strongly disordered NbN thin films with (a) $k_Fl$~0.42, (b) $k_Fl$~0.49, (c) $k_Fl$~0.82 and (a) $k_Fl$~1.23. The samples with $k_Fl$<1 show a pronounced peak in $\rho$-$H$. (e)-(f) Expanded view of ($\rho(H)$-$\rho(H=0)$) *vs.* magnetic field close to temperatures where the peak in the MR vanishes.

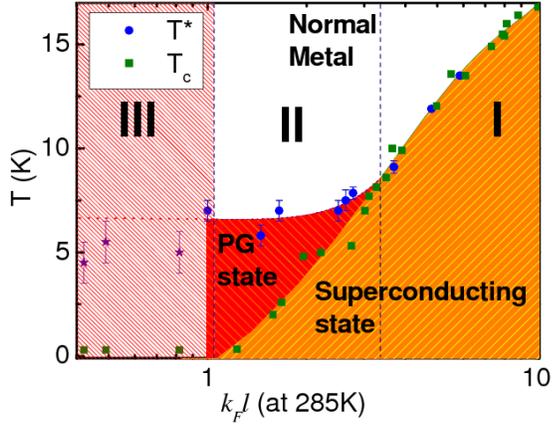

**Figure 7.** Phase diagram of strongly disordered NbN, showing $T_c$ (■) and $T^*$ (●) as a function of $k_Fl$. $T_c$ is obtained from transport measurements while $T^*$ is the crossover temperature at which the low bias feature disappears from the observed tunneling conductance. The samples with $k_Fl<1$ remain non-superconducting down to 300 mK. The three regimes with increasing disorder are shown as I, II, and III. A pseudogap (PG) state emerges between $T_c$ and $T^*$ for samples with $T_c \lesssim 6K$ (Regime II). The temperature at which the peak in the MR vanishes for the strongly disordered samples (Regime III) is also shown (★).

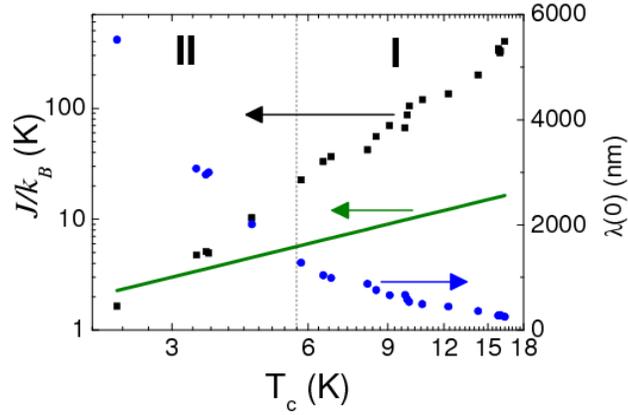

**Figure 8.** Superfluid stiffness ($J/k_B$) and penetration depth ($\lambda(T\rightarrow 0)$) for NbN films with different $T_c$. The solid line corresponds to $J/k_B=T_c$. Regime I and regime II corresponding to the phase diagram is delineated by the dashed vertical line.

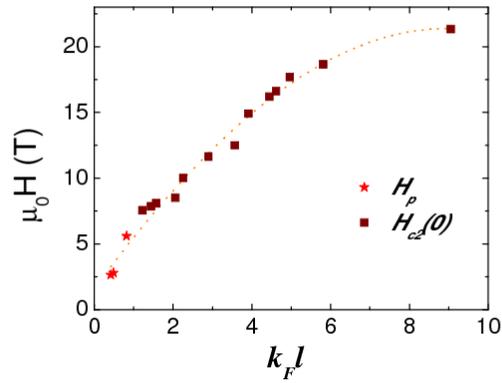

**Figure 9.** Variation of $H_{c2}(0)$ (for $k_Fl>1$) and $H_p$ (for $k_Fl<1$) as a function $k_Fl$.